\documentstyle[preprint,aps]{revtex}
\tightenlines

\def\slash#1{\raise.18ex\hbox{/}\kern-.50em #1}

\begin{document}

\def\spose#1{\hbox to 0pt{#1\hss}}
\def\ltapprox{\mathrel{\spose{\lower 3pt\hbox{$\mathchar"218$}}
 \raise 2.0pt\hbox{$\mathchar"13C$}}}
\def\gtapprox{\mathrel{\spose{\lower 3pt\hbox{$\mathchar"218$}}
 \raise 2.0pt\hbox{$\mathchar"13E$}}}
\def\inapprox{\mathrel{\spose{\lower 3pt\hbox{$\mathchar"218$}}
 \raise 2.0pt\hbox{$\mathchar"232$}}}


\draft
\preprint{DFTUZ/99/01}

\title{Correlation between UV and IR cutoffs in quantum field theory\\
       and large extra dimensions}

\author{J.L. Cort\'es\thanks{E-mail: cortes@leo.unizar.es}} 
\address{Departamento de F\'{\i}sica Te\'orica, Universidad de Zaragoza,
50009 Zaragoza, Spain}

\date{January 22, 1999}

\maketitle
\begin{abstract}
A recently conjectured relashionship between UV and IR cutoffs in an
effective field theory without quantum gravity is generalized in the
presence of large extra dimensions. Estimates for the corrections to
the usual calculation of observables within quantum field theory are
used to put very stringent limits, in some cases, on the characteristic 
scale of the additional compactified dimensions. Implications for the
cosmological constant problem are also discussed.   
 
\end{abstract}


\paragraph*{Introduction.}

The main subject of this note is to try to discuss the implications for 
low energy physics of the breakdown of effective field theory, with an
UV cutoff $\Lambda$, to describe a system in a finite volume
with a length $L$ when this length becomes smaller than a critical 
value which depends on the UV cutoff. One way to derive a limitation
of this type on conventional quantum field theory is based on the 
postulate~\cite{Bekenstein} that the maximum entropy in a box of
volume $L^3$ is proportional to the area of the box. A stronger 
constraint on the validity of quantum field theory is 
obtained~\cite{Kaplan} if one restricts the IR cutoff $L$ in order to
exclude states containing a black hole. Assuming a simple power 
dependence on the cutoffs for the corrections to a conventional
calculation in an infinite box without a UV cutoff one finds~\cite{Kaplan} 
that the uncertainty in the quantum field theory calculation is far
larger than what one naively would ascribe to gravitational effects.

The second ingredient of the present discussion is the very interesting
recent proposal of large compactified extra dimensions~\cite{Dimopoulos}
which allow to understand the observed weakness of gravity in a theory with
only one fundamental short distance scale, the weak scale.

When one considers simultaneously the possibility to have large
extra dimensions and also the limitations of quantum field theory
due to the impossibility to describe states containing a black hole
then it becomes important to try to estimate the errors in any 
conventional quantum field theory calculation which is the aim
of this short note.
 
\paragraph*{Deviations from QFT calculations.}

Let us assume, to start with, that the length $L$ of the finite 
box, which acts as an infrared cutoff, is smaller than the
compactification radius $R$ so we can consider a box of volume
$L^{3+n}$ in $3+n$ space dimensions. A restriction on the infrared 
cutoff can be obtained~\cite{Kaplan} by requiring to any state 
to have a Schwarzschild radius $L_0$ smaller than its size $L$.
If one denotes by $M$ the mass scale associated to the gravitational
constant in $4+n$ dimensions ($G = 1/M^{n+2}$) and uses the bound
$\rho \leq \Lambda^{4}$ for the energy density in the presence of
an UV cutoff $\Lambda$ then one has that the largest possible
value for $L_0$ is related to the UV cutoff through 
$L^{1+n}_0 \sim M^{-(2+n)}L^{3+n}\Lambda^{4+n}$. Then the 
condition $L_0 < L$ leads to a restriction combining the IR and 
UV cutoffs
\begin{equation}
L^{2}\Lambda^{4+n} < M^{2+n}
\end{equation}
which is the generalization of the constraint $L\Lambda^{2}<M_{P}$
obtained~\cite{Kaplan} in $3+1$ dimensions.

In order to estimate the corrections to a conventional quantum
field theory calculation done in an infinite box without a UV
cutoff one assumes~\cite{Kaplan} that they can be given as an
expansion in powers of $1/\Lambda$ and $1/L$. One can consider
two general classes of observables, \lq\lq chirally''
protected ($\hat{O}$) involving corrections with even powers
of the cutoffs exclusively and the remaining ones ($O$) which
have all power corrections. The result for an observable 
$\hat{O}$ including one loop radiative corrections will be

\begin{equation}
\hat{O}\approx \hat{O}_{QFT} \left[ 1 + \delta \hat{O}\right]
\end{equation}
where $\hat{O}_{QFT}$ is the calculation in QFT including one
loop radiative corrections and the correction $\delta\hat{O}$
can be estimated to be

\begin{equation}
\delta\hat{O} \sim {\alpha \over \pi} \left[
{E^{2}\over {\Lambda^{2}}} + {1\over {L^{2}E^{2}}}\right]
\geq {\alpha\over \pi}\left[
{E^{2}\over {\Lambda^{2}}} + {\Lambda^{4+n}\over {E^{2}M^{2+n}}}\right]
\end{equation}
where $\alpha$ is the coupling of the perturbative expansion and $E$
is the characteristic energy of the observable ${\hat O}$.

From this estimate for the correction to the field theoretical calculation 
it is possible to determine the choice for the UV cutoff which minimizes
the discrepancy in the calculation of ${\hat O}$ in QFT. The best choice
for $\Lambda$ corresponds to $\Lambda^{6+n}\sim E^{4}M^{2+n}$ and the 
minimal uncertainty in the calculation is

\begin{equation}
\delta_{min}{\hat O} \sim {\alpha\over \pi} 
\left({E\over M}\right)^{{4+2n}\over {6+n}}
\label{min}
\end{equation}
In the case of two extra dimensions one has $\delta_{min}{\hat O} \sim
(\alpha/\pi)(E/M)$ which is larger than the corrections that one would 
expect in an effective field theory calculation valid 
up to the scale $M$ which signals the onset of gravitational effects, 
$\delta {\hat O} \sim (\alpha/\pi)(E/M)^2$. 

If one considers an observable with no \lq\lq chiral'' protection the 
discussion can be repeated and the final answer is

\begin{equation}
\delta_{min}O \sim {\alpha\over \pi}
\left({E\over M}\right)^{{2+n}\over {6+n}}
\end{equation}
which for the case of two extra dimensions gives a correction proportional
to $\sqrt{E/M}$ to the QFT calculation.

From the success of the QFT one can derive a lower bound on the scale
$M$ of the gravitational coupling in $n+4$ dimensions. The bound will
be higher as the deviations of the experimental determination  from the
QFT calculation is smaller and for a given accuracy one will get more
stringent bounds as the energy increases.  
As an explicit example of a chirally-protected observable, which is in 
fact the most important quantitatively,
one can consider the anomalous magnetic moment of the electron, $g-2$,
which is known to an accuracy of $10^{-11}$. In this case one has a 
characteristic energy scale $E=m_e$ and demanding  $\delta_{min}(g-2)$
to be smaller than the experimental error leads, in the case of two
extra dimensions, to the bound $M > 100\,TeV$.  
This bound is more stringent that the bound obtained previously by
considering contributions of Kaluza-Klein (KK) modes. The very
weak coupling of KK-gravitons is compensated by their very large
multiplicity~\cite{Shrock} and one finds~\cite{Dimopoulos2,Shrock} a cross
section for emission of KK-gravitons of the order of $E^{2}/M^{4}$ where
$E$ is the center-of-mass energy and a corresponding lower bound 
$M > 30\,TeV$. 

If one considers the anomalous magnetic moment of the 
muon one has $E=m_{\mu}$ and then larger corrections than in the case 
of the electron but also the experimental error is larger ($\sim 10^{-8}$)
so that the corresponding bound for $M$ is lower. The same happens for
other observables at higher energies which are not determined 
experimentally with enough precission to give a bound comparable with 
the one obtained from $(g-2)_e$. 

If one considers a higher number of extra dimensions then the exponent
of $E/M$ in the deviations from QFT is bigger and then one finds 
smaller values for the lower bound on the graviational mass scale $M$.
In the particular case of six extra dimensions one finds 
$M\gtapprox 1\,TeV$ which is closer to the Fermi scale.

\paragraph*{Consistency of the determination of $\delta_{min}{\hat O}$.}

In the discussion of deviations from the QFT conventional calculation
we have assumed that the compactification radius $R$ in the extra 
dimensions is smaller than the lenght $L$ of the box. 
The relation~\cite{Dimopoulos,Shrock} between the $4+n$-dimensional 
gravitational coupling and the effective four-dimensional Newton constant,
$R^{n}M^{2+n}=M^{2}_{P}$ where $M_{P}$ is the Planck mass scale, can be
used to determine the radius $R$ in terms of the scale $M$,

\begin{equation}
R^{2} = {1\over M^{2}} \left({M_{P}\over M}\right)^{4/n}
\end{equation}

The result(\ref{min}) for the minimal deviation from the QFT calculation 
is obtained with an UV cutoff $\Lambda$ such that 
$\Lambda^{6+n}\sim E^{4}M^{2+n}$ and an IR cutoff $L$ such that
$L^{2}\sim M^{2+n}/\Lambda^{4+n}$. Then one has

\begin{equation}
L^{2} \sim {1\over M^{2}} \left({M\over E}\right)^{{{16+4n}\over {6+n}}}
\end{equation}

The condition $L<R$ gives a restriction on the energy scale $E$,

\begin{equation}
E > M \left({M\over M_P}\right)^{{{6+n}\over {n(4+n)}}}
\label{E}
\end{equation}   
which in the case $n=2$ can be rewritten as

\begin{equation}
E \gtapprox {m_{e}\over 10} \left({M\over {100\,TeV}}\right)^{5/3}
\end{equation}

Then in the case of two large extra dimensions with $M\sim 100\,TeV$
we have that the estimate (\ref{min}) for the deviations from QFT is 
valid for all proccesses with an energy scale $E\gtapprox 100\,KeV$. This
includes all the applications of relativistic quantum field theory. 
        
If one considers more than two extra dimensions then the range of energies
(\ref{E}) where the estimate (\ref{min}) for the deviations from QFT is
valid is reduced. In the particular case of six extra dimensions one has

\begin{equation}
E > M \left({M\over M_P}\right)^{1/5} \sim 1\,GeV 
\left({M\over {1\,TeV}}\right)^{6/5} \,,
\end{equation}
which in this case excludes, for $M$ close to the Fermi scale,
all the applications of QFT at energies $E\ltapprox 1\,GeV$. This
includes the evaluation of $(g-2)$ for the electron where $E = m_e$.
At this very low energy the IR cutoff which minimizes the 
deviation from the QFT evaluation is such that $L\gg R$ and then one 
can neglect the extra dimensions in the estimate of the deviations 
from QFT. In this case one has to replace $n$ by $n_{eff}=0$ and 
$M$ by $M_{eff}= M_P$ in (\ref{min}) and one has~\cite{Kaplan}

\begin{equation}
\delta_{min}[(g-2)_e] \sim {\alpha\over \pi} 
\left({m_{e}\over M_{P}}\right)^{2/3} 
\sim {\alpha\over \pi} \times 10^{-15} \,,
\end{equation}
which is very small compared with the experimental error in the 
determination of $(g-2)_e$. The stronger constraint on the 
scale $M$ will come, in this case, from high energy observables 
$E\gtapprox 1\,GeV$, where one expects deviations from QFT of the 
order of
\begin{equation}
\delta_{min}{\hat O} \sim {\alpha\over \pi} \left({E\over M}\right)^{4/3}
\,.
\end{equation}
   
\paragraph*{Cosmological constant.} 

We end up with a comment on the result of a QFT evaluation of the vacuum
energy density $\rho_{0}$ along the same lines. The result in perturbation
theory will be $\rho_{0}\sim \Lambda^{4}$ and in order to reproduce the
value suggested by the supernovae data an UV cutoff  
$\Lambda \sim 2.5 \times 10^{-3} eV$ is required~\cite{Shrock}. If one 
uses the relation $R = M_{P}/M^2$ between the compactification radius $R$ 
and the scale $M$ and the relation between the cutoffs leading to 
$\delta_{min}{\hat O}$ in the case of two extra dimensions 
($L\Lambda^{3}\sim M^2$) then one finds for the IR cutoff

\begin{equation}
L \sim 10^{36} \left({M\over {100\,TeV}}\right)^{4} R 
\end{equation}
which is much bigger than $R$. In this case one would expect that 
the extra dimensions can be neglected and one should repeat all
the considerations based on a correlation between IR and UV cutoffs
but now in four effective space-time dimensions. In this case the 
UV cutoff which minimizes the deviations from a QFT calculation is
$\Lambda\sim (E^{2}M_P)^{1/3}$; then in order to reproduce the
value of the UV cutoff, $\Lambda \sim 2.5 \times 10^{-3} eV$,  and then 
the smallness of the cosmological constant, one has to find a characteristic 
energy scale in the evaluation of the vacuum energy density 
$E_{0}\sim 10^{-18} eV$ (!). It is not clear what can be the origin of 
such a small energy scale.
Alternatively one can estimate the corresponding value of the effective 
IR cutoff through $L_{eff}\Lambda^{2}\sim M_P$ which gives 
$L_{eff}\sim 10^{28} cm$, a value comparable to the present horizon size. 
In order to predict what the cosmological constant should be one would
have to go beyond QFT to find the origin of these scales.

\medskip

I am grateful to M. Asorey, F. Falceto and A. Segui for discussions
and J.L. Alonso for reading the manuscript.
This work was supported by CICYT (Spain) project AEN-97-1680.

\end{document}